\documentstyle[preprint,aps,epsfig]{revtex}
\tighten
\begin{document}

\draft
\title{Resonance Structure in the Li$^-$ Photodetachment Cross 
Section}

\author{U. Berzinsh, \cite{byline}
G. Haeffler, D. Hanstorp,
A Klinkm\"uller, E. Lindroth\cite{byline2}, U. Ljungblad and D. J. 
Pegg$^{\ddagger }$}
\address{Department of Physics, Chalmers University of 
Technology\\
and G\"{o}teborg University, S-412 96 G\"{o}teborg, Sweden}

\date{\today}
\maketitle

\maketitle

\begin{abstract}
We report on the first observation of resonance 
structure in the total cross 
section for the photodetachment of Li$^-$. The structure arises 
from the 
autodetaching decay of doubly excited $^1P^o$  states  of Li$^-$ 
that are bound with respect to the $3p$ state of the Li atom. 
Calculations have 
been performed for both Li$^-$ and H$^-$ to assist in the 
identification of these resonances. The lowest lying 
resonance is analogous to the previously observed 
symmetrically excited 
intrashell resonance in H$^-$ but it is much broader. Higher lying 
resonant states are observed to converge on the Li($3p$)limit. These 
Rydberg-like resonances are much narrower and correspond to 
asymmetrically excited 
intershell states.
\end{abstract} 

\widetext

\narrowtext

  Processes such as double excitation and ionization of few-electron 
atomic 
systems provide valuable insights into the dynamics of the 
electron-electron 
correlation~\cite{Fan_83}. Double excitation has been 
studied 
in experiments involving photon-, electron- and heavy particle 
impact on 
atoms and ions. Photoexcitation, although restricted by the 
electric dipol 
selection rules, provides the potential for the highest energy 
resolution 
which is of considerable importance in the investigation of narrow 
and 
closely spaced resonance structures. In these investigations the 
helium
atom has been the most studied system~\cite{Mad_63,DXP_91}.
Negative ions, however, are 
 more sensitive to 
correlation effects than 
corresponding isoelectronic atoms or positive ions since for this 
member of 
a sequence the core field is weakest and therefore the masking of 
the 
subtler interelectronic interaction is reduced. The photo-double 
excitation 
of a negative ion involves the simultaneous promotion of a pair 
of electrons 
to relatively large distances from a singly-charged core. 
Under this condition 
the interaction between the pair of electrons can become 
comparable in strength to the weakened interaction of each 
electron with 
the core. The motion of the excited electrons then becomes highly 
correlated 
and the independent electron model ceases to be a valid 
approximation.

Until recently, most investigations of 
photo-double excitation in negative ions involved the H$^-$ ion.
Experimental studies of  
photodetachment in  this prototype anion
have been reported by Bryant and co-workers 
\cite{Ham_79,Har_90} and 
several computational techniques 
\cite{Her_75,Lin_86,Sad_92,Ho_92,Tan_94}
have been applied to 
account for the 
observed resonance structure. Negative ions of
heavier alkali metals has also been studied\cite{Pat_74}.

In this paper we report on the first observation of resonance 
structure in the total cross 
section for the photodetachment of Li$^-$. The 
resonances are of 
$^1P^o$ final state symmetry and are optically coupled to the 
ground state. 
 They arise from the photoexcitation and 
subsequent 
autodetaching decay of doubly-excited states of Li$^-$ that are 
embedded in 
continua representing an excited Li atom and a free electron. 
The resonances were observed to lie in the 
energy region between the $Li(3s)$ and $Li(3p)$ thresholds. 
Recently Pan, Starace and Greene used an eigenchannel R-matrix method to
predict the shape of the photodetachment cross section in the vicinity
of the Li$(n=3)$ threshold \cite{Pan} and the Li$(n=4,5,6)$ thresholds
\cite{Pan_94}.
The measurements reported here present us with an opportunity 
to compare experimental data on Li$^-$ with corresponding data on 
H$^-$~\cite{Ham_79}. 
Differences in the spectra of Li$^-$ and H$^-$ associated with the 
lifting 
of the degeneracy characteristic of the H-atom are apparent as are 
certain 
similarities.
We present also new calculations on Li$^-$ and H$^-$ which are used 
to explain the origin of the observed resonances.
The method accounts for full correlation between the outer
electrons and 
the autodetaching decay of doubly excited states is treated using
complex rotation.

Experimental investigations of narrow and relatively weak 
resonance structure in total photodetachment cross sections demand 
measurements of both high sensitivity and resolution. These 
conditions were achieved in the present experiment by the use of a 
collinear beam apparatus, described in detail elsewhere 
\cite{Han_92_1}, in which a tenuous beam of Li$^-$ is superposed 
with a laser beam over a common path of about 50 cm (see Fig.1). 

The energy resolution of our apparatus has in a recent 
experiment \cite{Ber_94} 
been shown to be limited
by the linewidth of the laser. In the present experiment the
 laser linewidth  
corresponds to an energy resolution 
of 25\thinspace $\mu$eV.
\begin{figure}
\epsfig{file=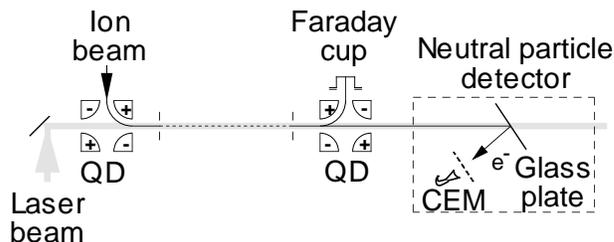, width=0.5\textwidth}
\caption {Portion of the collinear laser ion beam apparatus
showing the interaction and detection regions. Two quadrupole
deflectors (Q.D.) are used to merge the two beams.}
\label {fig1}  
\end{figure}

Li$^-$-ions were generated in a hot-plasma type ion 
source and accelerated to 4\thinspace keV. 
The tunable ultraviolet radiation used to transiently produce the 
doubly excited state by photoabsorption from the Li$^-$ ground 
state was generated by frequency doubling of the fundamental output 
of a XeCl excimer-pumped dye laser. 
Neutral Li atoms produced by photodetachment were 
separated from Li$^-$ ions by means of an electrostatic quadrupole 
deflector (Q.D.) and detected. The neutral atom detector consisted 
of a 
glass plate (coated with a thin layer of 
$In_2O_3:Sn$
to prevent charge build up) and a channel electron multiplier (CEM) 
that was employed to count the secondary electrons produced by 
the impact of the neutrals on the plate.
\begin{figure}
\epsfig{file=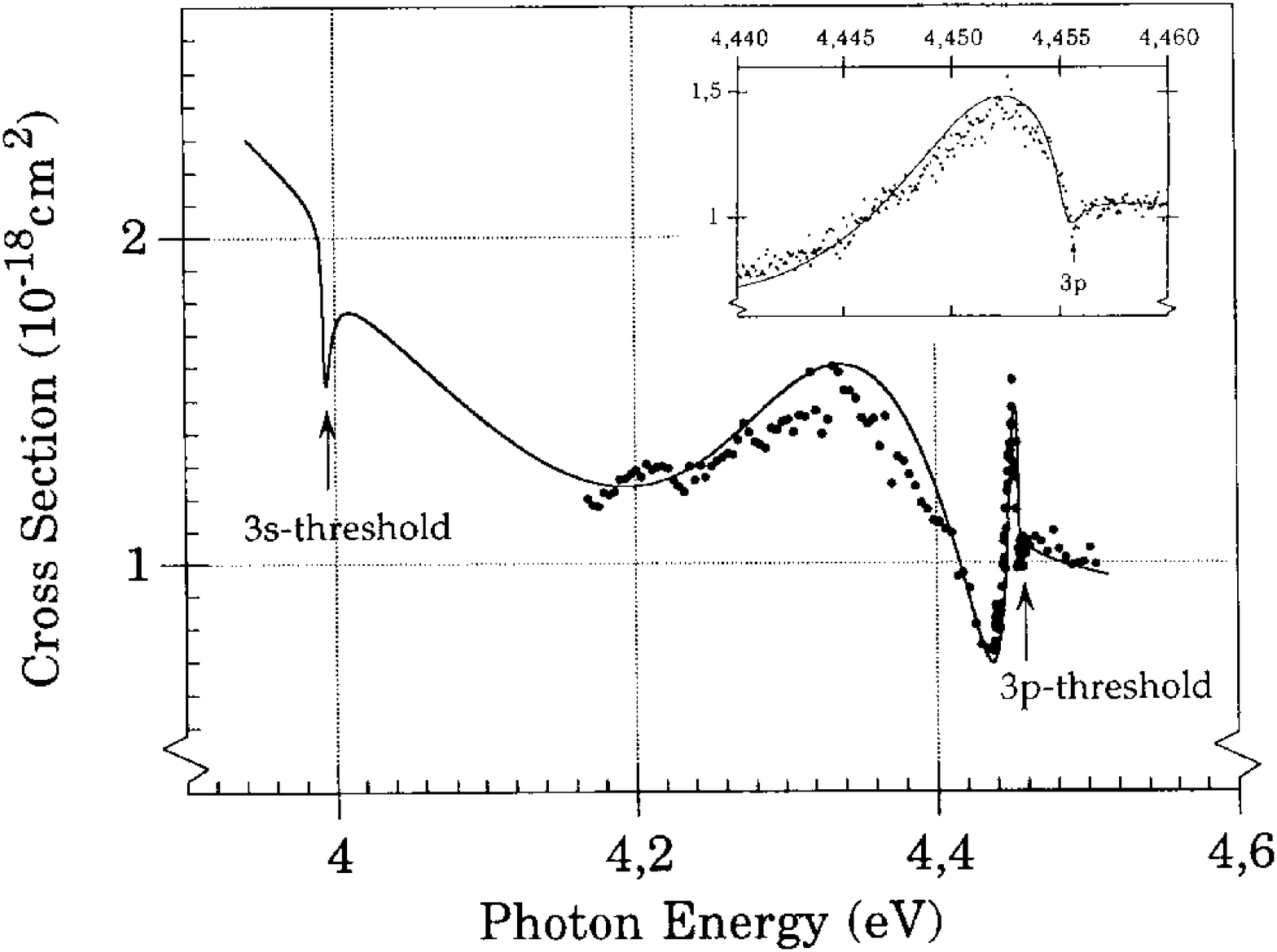, width=0.5\textwidth}
\caption{Total photodetachment cross section for Li$^-$ in the region
between the $3s$ and $3p$ thresholds. The dots  
represent the experimental data while the solid line
represents the theoretical result. The
inset shows an expansion of the region around the $3p$-threshold.}
\label{fig2}
\end{figure}
Two sources of background 
became evident in the experiment.
A contribution from neutrals 
produced by the detachment in collisions with the residual gas in 
the experimental chamber was rendered essentially negligible by 
maintaining a vacuum of about 
$10^{-9}$\thinspace mbar. A potentially more serious background 
source, however, was associated with photoelectrons generated on 
the glass plate of the neutral atom detector by the pulsed uv 
radiation. This contribution was greatly reduced by modulating the 
bias on a grid placed between the glass plate and the CEM. For a 
short time after the arrival of the laser pulse, a grid voltage of -
\thinspace40\thinspace eV was used to prevent the background 
photoelectrons from reaching the CEM. The grid voltage was 
reduced to zero about 0.5 \thinspace $\mu$s  after the laser pulse 
(the shortest flight time of the neutrals from the interaction 
region to the glass plate) in order to detect the neutrals 
constituting the signal. This gated detection procedure 
significantly enhanced the sensitivity of the measurement. The 
neutral atom signal, normalized to the intensities of the ion and 
photon beams, was used as a monitor of the relative total 
photdetachment cross section. 

The experimentally determined total photodetachment cross 
section is shown as the dots in Fig.2. All data points have been 
normalized to the theoretically calculated value at the $3p$ 
threshold. The scatter in the data is primarily caused by the 
small but unavoidable change in the overlap of the ion and laser 
beams as the wavelength of the laser was scanned. The 
normalization procedure 
was unable to 
account 
for the beam overlap variation. We estimated, however, that this 
effect causes a maximum error in the 
relative cross section over the whole spectrum 
of less than 10\%. The statistical scatter, mainly caused by the 
shot
noise in counting of the neutral particles, is less than 3\%.
The experimental data exhibits three significant features. There is 
a small 
and narrow dip in the cross section just at the $3p$ threshold which 
we interpret as the Wigner cusp arising from the opening of a new 
photodetachment channel. Second, there is a narrow resonance 
structure just 
below the threshold and, finally, a very broad resonance occupies a 
large part of the region between the Li($3s$) and Li($3p$) thresholds.

In order to explain the origin of the resonances in the 
observed cross section, we have calculated the 
expected spectrum of resonances in the photodetachment 
cross section of Li$^-$.
We have also made a calculation of H$^-$ for the purpose
of identifying differences associated with the presence or absence 
of a finite core. 
The calculated cross sections for Li$^-$ and H$^-$ are shown 
in Fig.2 and Fig.3, respectively, and the 
parameters 
associated with the intrashell, $(n=N)$, resonance (the broad resonance) 
in each spectrum are listed in 
Table I.
The calculation exploits a recently developed approach 
\cite{Lin_94}
which combines complex rotation, where the
radial coordinates are scaled with a
complex constant $r \rightarrow re^{i\theta}$,
 with the use of a discrete numerical 
basis set \cite{Sal_89}.
\widetext
\squeezetable
\begin{table}\label{tab:compare}
\protect\caption{Comparison between experimental and
theoretical results for intrashell $^1P^o$ states
of H$^-$ below the H$(n=3)$ threshold
and of Li$^-$ below the Li$(3p)$ threshold.
Energies and widths for the doubly excited states are given in eV.
Both transition energies (relative to the ground state) and binding energies
(reltive to the double detachment limit) are given.
( $^{a}$ Hylleraas wave functions and complex rotation
\protect\cite{Ho_92};  $^{b}$ partial waves included up to
l$_{max}$=3; 
$^{c}$ Hamm et al \protect\cite{Ham_79}.)}

\begin{tabular}{lccccccc}

&\multicolumn{2}{c}{\em Ho$^{a}$}
&\multicolumn{3}{c}{\em Present$^{b}$}
&\multicolumn{2}{c}{\em Experiment$^{c}$}
\\
&{\em E$_{bind.}$} &{\em $\Gamma$}  &{\em E$_{bind.}$} &{\em
$\Gamma$}
&{\em E$_{tran.}$}
&{\em E$_{tran}.$}
&{\em $\Gamma$}
 \\
 \tableline \\

 H$^-$
 & -1.705689 & 0.03240
 &\dec -1.706 &\dec 0.033
 &\dec 12.647 & 12.650$\pm$0.004
 & 0.0275$\pm$0.008
 \\
 Li$^-$
 & &
 &\dec -1.69 &\dec 0.36
 &\dec 4.32

 \end{tabular}
 \end{table}
The ground states of Li$^-$ and H$^-$ were  calculated 
by use of
perturbation theory while a representation  of the
spectrum of final states was obtained by diagonalization of the
 matrix for the
interaction between the two outer electrons. 
 This procedure yields a two-particle basis set which includes both doubly 
excited states, coupled to the continuum, and bound-continuum 
states.
 To construct the matrix we have typically
used 1400 basis set combinations,
including both
continuum and bound orbitals, but the selection was  limited
to $sp$, $pd$, and $df$ symmetries.  
The number of partial waves included is the only important 
approximation for H$^-$ and determines the number of digits given 
in
Table I. For Li$^-$, approximations are also made concerning the 
interaction
between the core and the outer electrons.
Our 
one-particle basis set consists
of Hartree-Fock orbitals obtained in the potential 
arising from the $1s^2$ core plus a non-local polarization
potential which accounts for the dominating correlation effects in 
neutral Li.
 Except for this
correction the core is assumed to be inert. 
The eigenvectors of the matrix 
are denoted $\Psi_n$ and they constitute a discretized description of 
the 
correlated final
states.  Some of these states are rather localized doubly excited 
states and
the real and imaginary parts of their complex eigenvalues 
correspond to
their energy, $E_r$, and halfwidth, $\Gamma /2$, respectively. 
 The remaining eigenvectors constitute a
discretized description of the bound-continuum channels.

The cross section for photodetachment in the complex rotation 
scheme has
been discussed  by Rescigno and McKoy \cite{Res_75}
and is calculated as
\begin{equation}
\label{eq:sigma}
\sigma(\omega)= \frac{e^2}{4\pi \varepsilon_0} 
\frac{4\pi}{3}\frac{\omega}{c}  
\Im( \sum_n \frac{ < \Psi_0
 \mid \sum_{j}{\bf r}_j e^{i\theta} \mid \Psi_n >
< \Psi_n \mid \sum_{j} {\bf r}_j e^{i\theta} \mid \Psi_0 >}
{E_n - E_0 - \hbar \omega} ),
\end{equation}
where each $\Psi_n$ represents a correlated  state with a 
complex energy, $E_n = E_r - i\Gamma/2$.
The initial ground state is denoted by
$\Psi_0$.  The usual expression which requires summation over 
discrete
final states, $\Psi_n$, and integration over continuum states is
replaced in this case by a summation over the discretized
two electron spectrum.  The use of complex rotation 
allows us to carry out the sum directly without any
special considerations close to the poles in the energy 
denominator.

The calculated cross section for
Li$^-$ obtained using Eq.~\ref{eq:sigma} is shown as the solid line in Fig. 
2. In general there is a good agreement with the 
experimental data.
The major discrepancy is in the height of the broad resonance, but this
 difference is within the estimated experimental error.
The resonance
 parameters, displayed in Table I, are obtained from the 
complex eigenvalues of the resonant states.
A simple connection between the resonance parameters and the
cross section
can be established if the form of the latter
was determined by the resonant state alone. The cross section could
then be
 approximated by the Fano profile~\cite{Fan_61}; 
\begin{equation}
\label{eq:Fano}
\sigma(\omega)= \sigma_0
\frac{\left(q + \varepsilon\right)^2}{1+\varepsilon^2}
\end{equation}
where 
$\varepsilon = 
\big[ \hbar \omega -\left(E_r - E_0\right) \big] / \left(\Gamma/2\right)$,
$q$ is a shape parameter and $\hbar\omega$
is the photon energy.
This expression is obtainable from  Eq.~\ref{eq:sigma} if the sum
over all states is replaced by one particular 
state.
 If the cross section 
is strongly influenced by more than one state, however, such a description 
will
certainly not be adequate
and as a consequence it will then no longer be possible
  to obtain the position and width of a particular
doubly excited state  from the total cross section.
This is indeed the situation in the studied region of the
Li$^-$  spectrum. 
 
The result
of the H$^-$ calculation shown in Fig. 3, is in close agreement with 
other calculations~\cite{Sad_92,Tan_94}.
The lowest energy
resonance is identified with a doubly excited state that has a 
calculated
energy of $12.647$ eV (relative the H$^-$ ground state)
and a width of $0.033$ eV.  These
values again agree well with other calculations, such as the 
accurate
results obtained by Ho~\cite{Ho_92} using Hylleraas-type 
wave functions.
It is also in agreement with the resonance parameters obtained by 
the analysis,
in form of a fit to a Fano line shape, of the experimental data by 
Hamm
et~al~\cite{Ham_79}. The comparison is displayed in Table I.  This 
resonant state is dominated by a few hydrogenic
configurations; $3s3p$($35$\%), $3p3d$($31$\%) and 
$3p4d$($11$\%). 
 
In Fig.2 it is seen that for Li$^-$ a broad resonance occupies 
essentially
 the whole region between the $Li(3s)$ and $Li(3p)$ thresholds.
 The calculated cross section is shown as a solid line and is in 
close
 agreement with a recent R-Matrix calculation~\cite{Pan}.
   The present
 calculation predicts a resonant state of width $0.36$ eV to lie 
$4.32$ eV
 above the ground state of Li$^-$.  This state appears to be 
analogous to the
 symmetrically excited intrashell
state in 
H$^-$. It has a very similar binding energy relative the double detachment 
limit, but is
 about one order of magnitude broader.  The broadening of the 
resonance in
 Li$^-$ arises from the strong coupling to the $3s \varepsilon p$
 continuum, which is not available below H($n=3$) in  
H$^-$. 
The resonant state in Li$^-$ is dominated 
 by the configurations $3p3d$ and $4s3p$ and 
 there appears to be no significant contributions to the 
 localized part of the wave
 function from configurations with one electron in the $3s$ orbital
 , which is also in contrast to the case of H$^-$.

While the H$^-$ intrashell resonance is well described by a Fano 
profile 
the width of the resonant state in Li$^-$ is too broad for this 
to be possible. The calculated width of $0.36$ eV is 
overlapping the $3p$ threshold  as well as the narrow resonance
seen just below it in Fig.2. The latter resonance is due to 
 asymmetrically excited Rydberg-like states.
The presence of the threshold and the interference between the  resonances 
 affects the shape of the cross section curve significantly.  The 
assumption
necessary to obtain a Fano profile, that only one state determines 
the shape 
is thus not valid.  The interference results in a narrowed
structure in the spectrum, especially on the high energy side, 
compared to that
which would arise from a hypothetically isolated doubly excited 
state of width
$0.36$ eV.
Rydberg-like resonant states are also apparent in the calculated 
spectrum of
H$^-$.  

These states are bound relative H$(n=3)$ by the strong dipolar field
between the two electrons arising due to the degeneracy of the
H$(3\ell)$ states and resulting in nearly equal admixtures of $3snp$ and
$3pns$ (or $3pns$ and $3dnp$) configurations their composition.  The
Rydberg-like states in Li$^-$, however, do not have this character.
They are completely dominated by $3pn\ell$ configurations, with $n \gg
3$, and the dipolar field is in this case insignificant. The explanation
for the existence of the Rydberg states in Li$^-$ is the inability of
the monopole part of the electron-electron interaction to screen the
singly-charged core completely. The residual nuclear attraction binds
the states below the $Li(3p)$ threshold.

To conclude, we have studied the total cross section for 
photodetaching an electron from  the Li$^-$ ion in the energy 
region between the $3s$ and $3p$ thresholds. This region is 
dominated by a 
broad resonance. A narrower resonance structure lies just 
below the $3p$ threshold . By comparing the data with 
calculated cross sections for 
Li$^-$ and H$^-$ we have been able to identify the broad resonance 
in the
Li$^-$ spectrum as being associated with the presence of a 
symmetrically excited intrashell doubly excited state, analogous to, but much 
broader than, the intrashell
state in H$^-$. The sharper 
resonance
structure is identified with asymmetrically excited Rydberg-like 
resonant states. Our intention is to extend the measurements to 
higher 
lying resonances  to further  investigate the similarities
and differences between the photodetachment spectra of Li$^-$ and 
H$^-$.

Financial support for this research 
was received from the Swedish Natural Science
Research Council (NFR). One of us (DJP) whish to acknowledge  
the support from the Swedish Institute and the US 
Department of Energy.



\end{document}